\documentclass{elsart}
\usepackage{natbib}
\usepackage{epsfig}
\usepackage{journals} % INAOE
\def\hii{H\,{\sc ii}}

\begin{document}
\runauthor{Legrand Fran\c{c}ois}
\begin{frontmatter}
\title{Slowly cooking galaxies \thanksref{X}\thanksref{iap}}
\author[INAOE]{Fran\c{c}ois Legrand}

\thanks[X]{I thanks R.J. Terlevich for suggesting the title of this paper}
\address[INAOE]{Instituto Nacional de Astrofisica, Optica y
  Electronica, Tonantzintla, Apartado Postal 51 y 216, 72000 Puebla,
  Pue. MEXICO}
\thanks[iap]{This work is part of a PhD Thesis done at IAP (Paris)
  under the supervision of D. Kunth.}
%\thanks[ads]{This research has made use of NASA's Astrophysics Data System Abstract Service.}
\begin{abstract}

Recent spectroscopic observations of IZw~18 have revealed
homogeneous abundance throughout the galaxy and several observations
of other starburst galaxies have shown no significant gradient or
discontinuity in the abundance distributions within the HII regions.
I thus concur with \cite{TT96} and \cite{DRD97} that these observed
abundance homogeneities cannot be produced by the material ejected
from the stars formed in the current burst and result from a
previous star formation episode. Metals ejected in the current burst
of star formation remain most probably hidden in a hot phase and are
undetectable using optical spectroscopy. Combining various
observational facts, for instance the faint star formation rate
observed in low surface brightness galaxies \citep{VZHSB97}, I
propose that a low and continuous star formation rate occurring during
quiescent phases between bursts is a non negligible source of new
elements in the interstellar medium. Using a spectrophotometric and
chemical evolution model for galaxies, I investigated the
star formation history IZw~18. I demonstrate that the continuous star
formation scenario reproduces all the observed parameters of
IZw~18. I discuss the consequences of such a quiet
star formation regime.

\end{abstract}
\end{frontmatter}

\typeout{SET RUN AUTHOR to \@runauthor}

\section{Introduction}

Understanding galaxies formation and evolution is one of the most
challenging issues of modern astrophysics. In this field, low-mass
dwarfs and irregular galaxies have progressively reached a particular
place. Indeed, in hierarchical clustering theories these galaxies are
the building blocks of larger systems by merging
\citep{KWG93,PWKO96,LKGGPFVIG97}. Moreover, as primeval galaxies may
undergo rapid and strong star formation events \citep{PP67}, nearby
dwarf starburst galaxies or Blue Compact Galaxies (BCDG) of low
metallicity can also be considered as their local counterparts. 
Therefore the study of low redshift starbursts is of major interest 
for our understanding of galaxies formation and evolution.

During a starburst, the massive stars produce and eject metal-rich gas
into the interstellar medium, but the timescale for chemical
enrichment is far from being constrained. Is the process so quick that the
newly synthesized elements are immediately detectable in \hii\ regions ?
Is there a time delay between the release of nucleosynthesis
products and the chemical pollution of the star-forming regions ?
Answering these questions is crucial for the interpretation of the
abundances measurements in star-forming galaxies and their chemical
evolution.

\cite{KS86} first proposed that metals produced in a burst of star
formation are likely to enrich very quickly the surrounding \hii\
region. If true, the present burst in IZw~18 could alone account for
its observed metallicity \citep{KMM95} and this would explain why no
galaxy with a metallicity lower than that of IZw~18 has ever been found
despite extensive searches \citep{T82,TMMMC91,MMCA94,ITL94,TST96}.

Recently, \cite{RK95} argued that the newly synthesized elements
cannot be dispersed over scales larger than a few hundred parsecs in a
timescale $\leq$ 100 Myr, predicting that abundance discontinuities
should be observed in young starburst galaxies between the central
\hii\ regions (``auto-enriched'' by the massive ionizing stars) and
more external regions relatively free of recent chemical
pollution. However, recent observations of IZw~18
\citep{LKRMHW99,VZWH98} revealed a homogeneous abundance throughout
the galaxy (HII and HI) and several studies of other starburst
galaxies \cite[][and references therein]{KS97b} have shown no
significant gradient or discontinuity in 
the abundance distributions within the HII regions. This suggest that
the metals produced in the current burst are invisible in the optical
and remain hidden in a hot X-rays emitting phase as discussed by
\cite{TT96,DRD97,KS97b,PIL99}. An important consequence of this
is that the observed metals in IZw~18 and other starburst come from
previous star formation event which nature have to be specified.

On the other hand it is easy to show that the current SFR in starburst
galaxies cannot be maintained during a long time without consuming
most of the gas and producing excessive enrichment. It is thus
generally assumed that the star formation history of these objects is
made of a succession of burst separated by rather long quiescent
periods during which they are likely to appear as low surface
brightness or quiescent dwarf galaxies. However, even among these
objects none has been found with 
a star formation equal to zero \citep{VZHSB97}. All of them present
very low but non zero star formation rate (SFR). Indeed this is a strong
indication that star formation at a very
low level occurs even between bursts and that the metallicity still
increases slightly during 
these periods. This led \cite{LKRMHW99} to propose the existence of a
small but rather continuous SFR during the lifetime of galaxies and
suggest that this regime of star formation can be alone responsible of
the observed metals in IZw~18. Preliminary results \citep{LK98} seem
to agree with this hypothesis.

I will present here new results I obtained \citep[detailed
calculations can be found in ][]{L99} using a
spectrophotometric and chemical evolution model in order to
constrain the past star formation history of IZw~18. Particularly, I
will show how the continuous low star formation regime proposed by
\cite{LKRMHW99} can account for the observed metallicity in
IZw~18. Finally, I will discuss the consequences of such a star
formation regime .

\section{Modeling the past star formation history of IZw~18}
\subsection{The model}
In order to investigate the star formation history of
IZw~18, I used the spectrophotometric model coupled with the chemical
evolution program ``STARDUST'' described by \cite{DGS99}
The main features of the model can be found in \cite{L99}. I used 
a typical IMF  described as a power law in the mass range 0.1-120 $\rm
M_{\odot}$        
      \begin{equation}
      \phi(m)=a.m^{-x} 
      \end{equation}
      with constant index x of 1.35 \citep{S55}.\\
Two regimes of star formation has been investigated: 
       \begin{itemize}
         \item A continuous star formation during the lifetime of the
           galaxy. The SFR is low and
           directly proportional to the total mass of available gas.
         \item Bursts of star formation during which all the stars
           are formed in a rather short time. 
       \end{itemize}

The model provide us with both the abundances and the spectra at each time.

\subsection{Continuous SFR}

\begin{figure}
%\picplace{7cm}
\begin{center}
\fbox{ \psfig{figure=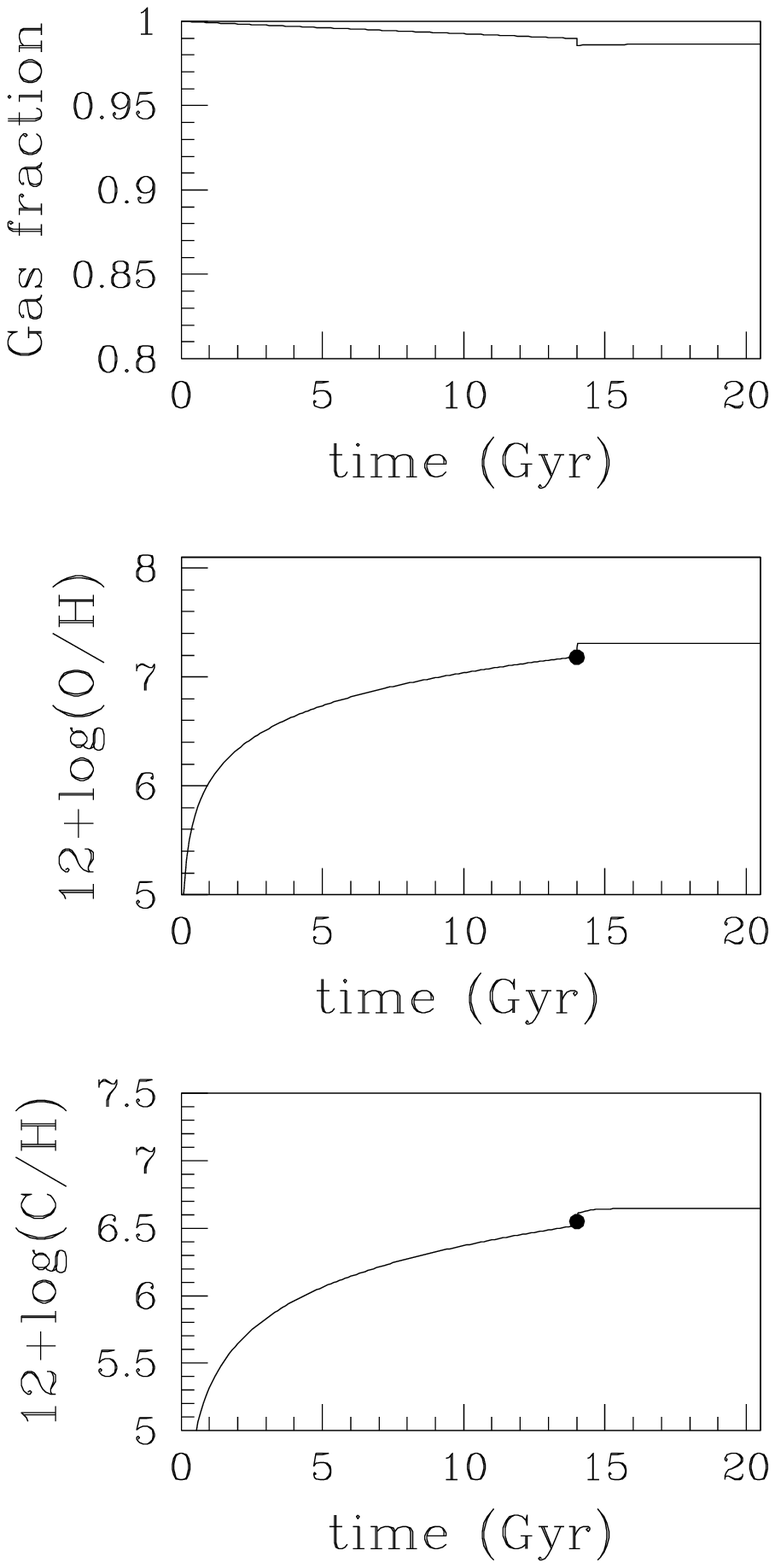,clip=,bbllx=15pt,bblly=150pt,bburx=300pt,
  bbury=700pt,height=8cm,angle=0} a } \fbox{ \psfig{figure=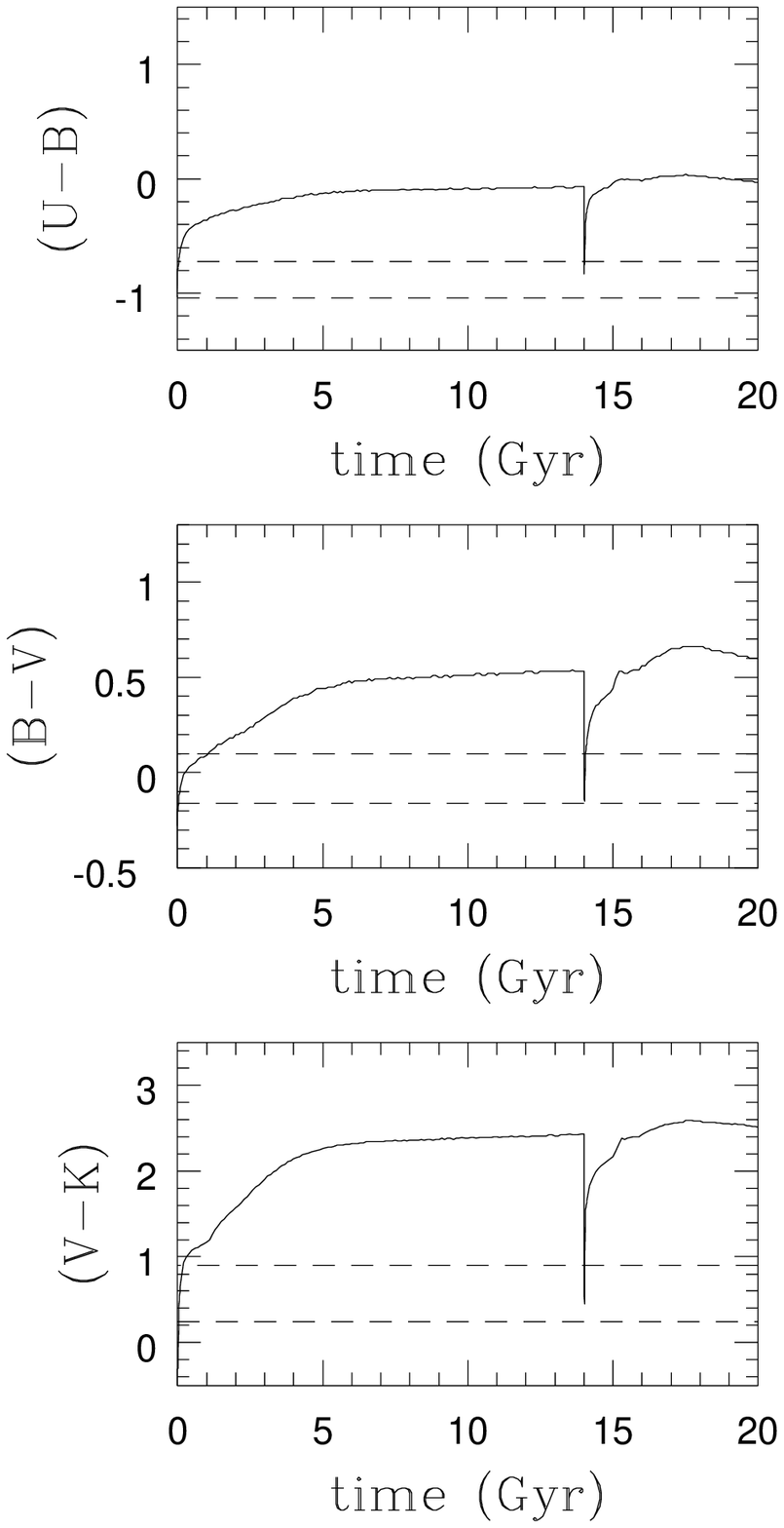,clip=,bbllx=15pt,bblly=150pt,bburx=300pt, bbury=700pt,height=8cm,angle=0} b}
\end{center}
\caption[]{a) Time evolution of the gas fraction, oxygen and carbon
  abundances for the continuous SFR model with a Salpeter IMF. The
  dots represent the measured abundances \citep[][for O and C
  respectively]{SK93,GSDS97}. b) Colors evolution (U-B, B-V, V-K). The
  dashed lines delimit the zone of compatibility between the model and
  the observations. The colors (corrected for the nebular
  contribution) are from \cite{S98} for (U-B) and (B-V) and from
  \cite{T83} for (V-K).} 
  \label{fig:cstSFR} 
\end{figure}

As all the galaxies containing gas are known to have a non zero SFR,
\cite{LKRMHW99} proposed the existence of a faint but continuous SFR during the
lifetime of the galaxies. In order to constrain this SFR I used
the model described before. Assuming that the present burst in IZw~18
is the first one in the history of this galaxy, but that this object
has undergone a faint but continuous SFR during its lifetime, I adjusted the
continuous SFR to reproduce the observed {\bf oxygen} abundance after
14~Gyrs. I found that a continuous SFR of $ 10^{-4}\,g\
M_{\odot}\,yr^{-1} $ where $g$ is the fraction of gas (in mass)
available in the galaxy, can reproduce the observed oxygen abundance
in IZw~18 after 14~Gyrs. In order to reproduce the present colors I
added a burst with the characteristics of the current one as given by
\cite{MHK98}. We have to keep in mind that the metals produced by this
burst are not yet visible so the metal measurements trace the
metallicity {\bf before} the burst.
The results of this model are presented in Fig.~\ref{fig:cstSFR}.

We can notice that within the error bars this model can reproduce all
the observations. The fraction of gas consumed remain very low thus
$g$ is always close to 1 and the SFR is rather constant.

%\begin{figure}
%\picplace{7cm}
%\psfig{figure=aazncc.vr.coul.ps,clip=,bbllx=40pt,bblly=330pt,bburx=590pt,
%  bbury=700pt,height=6.1cm,angle=0}
%\caption[]{Colors evolution (U-B, B-V, V-K) for the continuous SFR
%  model with a Salpeter IMF. The dashed lines delimit the zone of
%  compatibility between the model and the observations. The colors
%  (corrected for the nebular contribution) are
%  from \cite{S98} for (U-B) and (B-V) and from \cite{T83} for (V-K).}
%  \label{fig:cstSFRcolors} 
%\end{figure} 

\section{Generalization and Consequences}

Assuming that this continuous SFR occurs sporadically throughout the
galaxy, the homogeneity of the abundances (within the NW region but
also between NW and SE regions) is a natural outcome of this model;
the rather uniform spatial distribution of the formed stars and the
long time evolution ensuring a homogeneous mixing of the metals. The
physical process which could support such a extended star formation
have to be precised. Indeed, as in LSBG, the mean density seems to
remain under the critical threshold of instability for star formation
\citep{T64,C81,K89,VZHSB97}. However, the HI halo is certainly not
monolithic nor static but formed of many small clouds. When these
clouds collide, the
density should increase and may locally exceed the threshold. A study
of the processes which could be responsible of this star formation
regime is planed.

\begin{figure}
\begin{center}
\psfig{figure=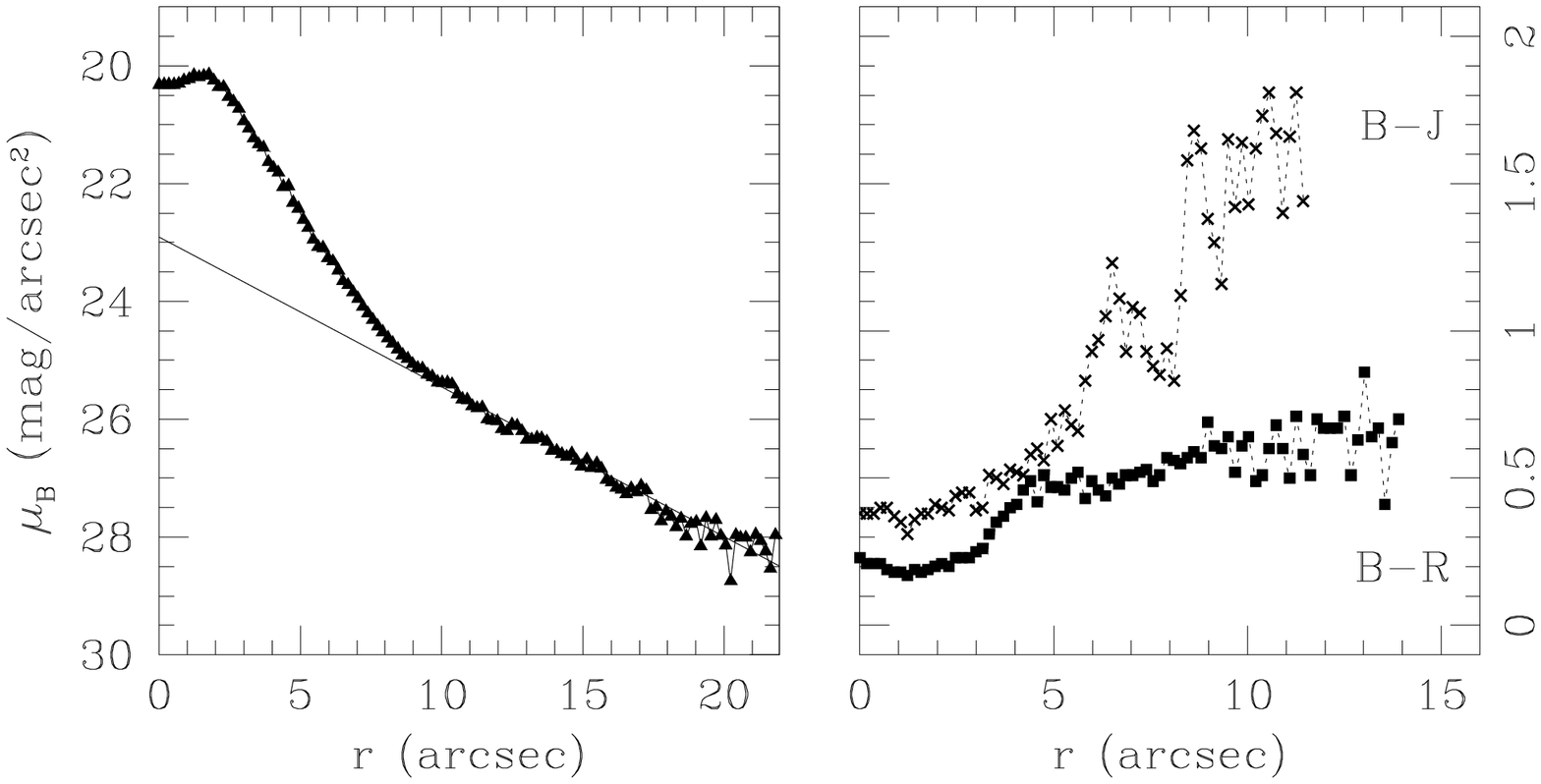,clip=,bbllx=10pt,bblly=240pt,bburx=310pt,
  bbury=550pt,height=7cm,angle=0}
\end{center}
  \caption[]{IZw~18 surface brightness profile \citep{KO99}}
 \label{fig:sbprof}
\end{figure}

Assuming that the continuous SFR occurs throughout the whole HI halo
of the galaxy ($60\times45''\ $) I predict a surface brightness of the
old underlying population of the order of 28 $\rm mag\,arcsec^{2}$
in V and 26 $\rm mag\,arcsec^{2}$ in K. These values are an upper
limit (in $\rm mag\,arcsec^{2}$); if a fraction of metals is ejected
out of the galaxy, the SFR 
needed to produce the observed abundances will be higher and the total
luminosity and surface brightness will be increased. Moreover the density limit
adopted for the continuous SFR is a lower limit and the region where
the continuous SFR can occur may be smaller, resulting in higher
surface brightness. However, the extreme
faintness of the old underlying population probably explains why no
strong evidence for its existence has been found in IZw~18
\citep{T83,HT95} until recently when reanalyzing HST archive images 
\cite{ATG99} found stars older than 1~Gyr. Moreover, preliminary surface
brightness profiles of IZw~18 (Fig.~\ref{fig:sbprof}) published by \cite{KO99} 
indicate a surface brightness of at least 28 $\rm mag\,arcsec^{2}$ in
B (may be lower) in the external parts of the galaxy (at 20'' from the
center). This results still have to be confirmed, but it agrees
with our predictions.

I also evaluated the number of massive stars ($M\,>\,8\,\rm
M_{\odot}$)  formed to be about 120 stars (an open cluster) every
140~Myrs. This not appears unrealistic.

I also compared this continuous SFR with the ones observed in LSBG and
quiescent dwarfs \citep{VZHS97a,VZHS97b,VZHSB97}. As these objects
have different masses, I normalized the SFR to the total HI mass
observed. It appears that the continuous SFR as predicted by our 
scenario is comparable, relative to the HI mass, to the lowest
SFR observed in quiescent and low surface brightness
galaxies \citep[see ][]{L99}. 

\begin{figure}
%\picplace{7cm}
\begin{center}
\psfig{figure=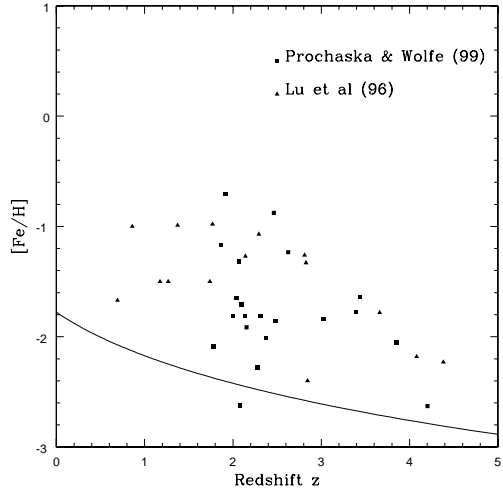,height=7cm,angle=0}\label{fig:compdla}
\end{center}
\caption[]{Comparison of the predicted and observed evolution with
  redshift of the abundance [Fe/H]. The points represent the data from
  \cite{LSBCV96} and \cite{PW99} and the solid line the model prediction
  for a constant star formation rate.}
\end{figure}

If a continuous star formation rate exists in IZw~18, it must
exist in other dwarf galaxies, and may be, in all
galaxies. If true, this explain why no galaxy with a metallicity lower
than that of IZw~18 has been found and why all the HI clouds detected by blind
surveys has all turned to be associated with stars \citep{B97}.
We can also expect that such a continuous low SFR
occurs in the outskirts of spirals, at few optical radius, where the
density is low. As a matter of fact, the extrapolation of the
abundance gradients in these objects lead to abundances comparable
to that of IZw~18 at radial distances of about three optical radii
\citep{FWG98a,HW99}. As this corresponds to the size of the halos or
disks susceptible to give rise to metallic absorption in quasar
spectra \citep{BB91}, we can also compare the time evolution of the
metallicity with the abundances measured in quasars absorption
systems. This comparison is done in Fig.~\ref{fig:compdla}. The
abundances predicted by the model mimic the lower envelope of these
measurements. If we assume that these absorption systems are
associated with galaxy halos \citep{LBTW95,TLS97}, this indicates that
such a process can account for a minimal enrichment of the ISM with time.

\section{Conclusions}

Various observations suggest that the metals produced by the massive
stars during a burst are not immediately visible using optical
spectroscopy. They should be in a hot phase emitting in the X-rays
range. Thus the observed metals has been produced during former star
formation event. Using the fact that we don't know any galaxy
containing gas with a SFR equal to zero, I propose the existence of a
low continuous SFR during the lifetime of galaxies. Using a
spectrophotometric model coupled to a chemical evolution model for
galaxies, I have shown that such a star formation regime is sufficient to
reproduce alone the observed metallicity of IZw~18 and can account for
various observational facts as the the presence of star formation in
quiescent dwarfs and LSBG, the apparent absence of galaxies with a
metallicity lower than that of IZw~18, the apparent absence of HI clouds
without optical counterparts, the homogeneity of the metal abundances in dwarfs
galaxies, the metal content extrapolations to the outskirts of spiral
galaxies and the metallicity increase with time in the most
underabundant quasars absorption systems. I thus conclude that, even
if starbursts 
are strong and important events in the life of galaxies, their more
subdued but continuous star formation regime cannot be ignored when
accounting for their chemical evolution.

\bibliography{/home/legrand/LATEX/BIBLIOGRAPHIE/bibliographie}

\end{document}